\documentstyle[prl,aps,multicol]{revtex}
\renewcommand{\narrowtext}{\begin{multicols}{2} \global\columnwidth20.5pc}
\renewcommand{\widetext}{\end{multicols} \global\columnwidth42.5pc}
\begin{document}%
\draft

\title{ Random Matrix Model and the
  Calogero-Sutherland Model: A Novel Current-Density Mapping}

\author{N. Taniguchi$^a$\cite{address-j}, B. S. Shastry$^b$ and B. L.
  Altshuler$^c$}

\address{$^a$Department of Physics, Massachusetts Institute of Technology,
  Cambridge, MA 02139, USA\\$^b$Indian Institute of Science, Bangalore -
  560 012, India\\ $^c$NEC Research Institute, 4 Independence Way,
  Princeton, NJ 08540, USA }


\maketitle

\begin{abstract}
  We investigate the relation between the invariant correlators of random
  matrix theory and correlators of the integrable one-dimensional systems.
  Starting from the relation between correlators for the coupling
  strengths $\lambda =1/ 2$, $1$, and $2$, we explore the local
  current-density mapping applicable to arbitrary $\lambda$ including {\em
    irrational\/} values, which results from the novel structure of
  the Calogero-Sutherland model. We find an interesting and novel
  relationship between equal time current and density correlations for any
  coupling, which exist {\em in addition} to the usual Ward Identities for
  this class of systems.
\end{abstract}


\narrowtext

Spectral statistics of complex quantum systems such as quantum dots and
quantum billiards exhibit striking universal behaviors irrespective of
their microscopic details.  These systems are characterized by the
repulsion of energy level.
To describe such spectral statistics in quantum
chaotic systems, random matrix theory (RMT) and Wigner-Dyson statistics
has been successfully applied~\cite{Haake,Mehta,Bohigas91}.
The universality classes of a given system is entirely determined by its
generic (spin and/or time-reversal) symmetry of the system: orthogonal
(spinless) and symplectic (with spin-orbit interaction) for T-invariant
systems, and unitary for T-breaking systems.
Typically the level repulsion is characterized by the level spacing
$\varepsilon$ distribution proportional to $\varepsilon^\beta$ for small
$\varepsilon$, where $\beta = 1$, $2$ and $4$ respectively for the
orthogonal, unitary and symplectic classes.
It is worth mentioning that that the universality of Wigner-Dyson
statistics corresponds to the {\em bulk} scaling limit of matrix models.
It is also possible to get new and different universality classes from
various edge scaling limits.

RMT is connected intimately with the Calogero-Sutherland model
(CSM)~\cite{Calogero69,Sutherland71,Moser75} which describes $N$ fermions
located on  a ring of the perimeter $L$.  The interaction between
these particles is pairwise and is inversely proportional to the square of
the arc distance:
\begin{equation}
  H={1 \over {2m}}\sum_i {p_i^2}+{{\hbar ^2\lambda (\lambda -1)}
    \over m}\sum_{i<j} {{{\phi ^2} \over {\sin ^2(\phi
        r_{ij})}}}-E_0,
\label{CSM}
\end{equation}
where $r_{ij}\equiv r_i-r_j$, $p_i=-i\hbar\partial / \partial r_i$, $\phi
=\pi / L$, $\lambda$ determines the sign and the strength of the
interaction, and $E_0=N(N^2-1)\phi ^2\lambda ^2/ 6m$. We use the unit
$\hbar =1$   hereafter.

The ground state wavefunction of Eq.~(\ref{CSM}) is given by a Jastrow
function, $\psi _0(r_1,\cdots, r_N)=\prod_{i<j} {\sin ^\lambda (\phi
  r_{ij})}$.
The Jastrow form of the wave function enables  ground state averages
(in the thermodynamic limit) to be identified with  an
average over Wigner-Dyson ensembles of random matrices for coupling
constants $\lambda = \beta/2 = 1/ 2$, $1$, and $2$.
In particular, the static density-density correlation function $\langle 0|
\rho(r) \rho(0) |0\rangle $ of CMS, where $\rho=\sum_i \delta(r-r_i)$ is
the density operator, can be connected straightforwardly to the two-point
correlator of density of states (DOS) $R_2(\omega)$ of RMT for these
values of $\lambda$~\cite{Sutherland92}.


Recently the connection between RMT and CSM has been extended from the
static correlations to the dynamical
correlations~\cite{Simons93,Beenakker93a,Narayan93}.
In the framework of RMT, dynamical correlations of CSM correspond to the
parametric correlations depending on the external parameter.
Take the random matrix which is perturbed by the external field,
$H(X)=H_0+X\Phi$, where $H_0$ is a random matrix belonging to one of the
Dyson ensembles, and $\Phi$ is a fixed traceless member of the same ensemble.
The parametric DOS correlator $K(\Omega, X) \equiv \langle \nu(\bar E,\bar
X) \nu(\bar E + \Omega,\bar X + X) \rangle$, where $\nu(E,X) = \sum_n
\delta(E-E_n(X))$ with $E_n(X)$ being the spectrum of $H(X)$ and $\langle
\cdots \rangle$ denotes a statistical average in some interval of energy
and/or external parameter, was
evaluated analytically by use of the supermatrix method~\cite{Efetov83}.
The resulting function $K(\Omega,X)$ is found to be universal after
rescaling $\omega \equiv \Omega / \Delta$ and $x^2 \equiv (X/\Delta)^2
\langle {\left( {\partial E_n(X)/ \partial X} \right)^2} \rangle$.

The connection with CSM is provided when we substitute $\pi\omega \to r$
and $\pi^2 x^2/2 \to -it$ in the expression of $K(\Omega,X)$, which
produces the dynamical density-density correlator of CMS $\langle
0|\rho(r,t)
\rho(0,0)|0\rangle$ for $\lambda = 1/2$, $1$, and $2$.
We remark that subsequently the dynamical density-density correlator of
CMS was evaluated for arbitrary rational values of $\lambda$ by use of Jack
polynomials~\cite{Ha94}.




In quantum dots or RMT, there are two kinds of two-point
correlation functions which become universal~\cite{Zirnbauer86b,Taniguchi95}.
In terms of the retarded and advanced Green functions, these two universal
functions are defined by
\widetext
\begin{eqnarray}
  &&k(\omega ,x)=-{1 \over 2}+{{\Delta ^2} \over {2\pi ^2}} \int
  {d\bbox{r}_1d\bbox{r}_2\,\langle G_{E+\Omega ,\bar
      X+X}^R(\bbox{r}_1,\bbox{r}_1)}G_{E,\bar
    X}^A(\bbox{r}_2,\bbox{r}_2)\rangle. \label{def:k}\\ &&n(\omega ,x) =
  {{\Delta ^2} \over {2\pi ^2}}\int {d\bbox{r}_1d\bbox{r}_2\,\langle
    G_{E+\Omega ,\bar X+X}^R(\bbox{r}_1,\bbox{r}_2)}G_{E,\bar
    X}^A(\bbox{r}_2,\bbox{r}_1) \rangle,\label{def:n}
\end{eqnarray}
\narrowtext\noindent
where $G_{E,X}^{R,A}(\bbox{r,}\bbox{r}')=\langle {\bbox{r}} |(E-H(X)\pm
i0)^{-1}| {\bbox{r}'} \rangle$  denotes the retarded and advanced Green
functions.
The DOS correlator is given by $K(\Omega,X) = \Delta^{-2} \mbox{Re} [1 +
k(\omega,x)]$, whereas $n(\omega,x)$ has determined the response
depending on the wavefunctions as well as on energy spectra.

For all the three universal classes (orthogonal, unitary, symplectic), the
analytical result for $k(\omega ,x)$ and $n(\omega ,x)$ were already
obtained~\cite{Simons93,Taniguchi95}. As a simplest case, these invariant
correlators for the unitary class is
given by
\widetext
\begin{mathletters}
\label{kn-unitary}
\begin{eqnarray}
  &&k_u(\omega ,x)={1 \over 2}\int_1^\infty {d\lambda _1\int_{-1}^1 {d\lambda
      \,e^{i\pi \omega (\lambda _1-\lambda )-\pi ^2x^2(\lambda
        _1^2-\lambda ^2)/ 2}}},\\
  &&n_u(\omega ,x)={1 \over 2}\int_1^\infty {d\lambda _1\int_{-1}^1 {d\lambda
      \,{{\lambda _1+\lambda } \over {\lambda _1-\lambda }}e^{i\pi \omega
        (\lambda _1-\lambda )-\pi ^2x^2(\lambda _1^2-\lambda ^2)/ 2}}},
\end{eqnarray}
\end{mathletters}
\narrowtext\noindent
%
%
{}From the analytical expressions for $k(\omega,x)$ and $n(\omega,x)$, it is
straightforward to observe that there is a simple differential relation
connecting between these functions for all three universality
classes~\cite{Taniguchi95}:
\begin{equation}
  2{\partial \over {\partial x^2}}k(\omega ,x)={{\partial ^2} \over
    {\partial \omega ^2}}n(\omega ,x).
\label{kn-relation}
\end{equation}
The identity (\ref{kn-relation}) allows us to understand what is the
counterpart of $n(\omega,x)$ in the context of CSM.
Indeed, by comparing Eq.~(\ref{kn-relation}) with the
continuity relation
\begin{eqnarray}
  &&{{\partial ^2} \over {\partial t_1\partial t_2}}\left\langle 0
\right|\rho (r_1,t_1)\rho (r_2,t_2)\left| 0 \right\rangle \nonumber\\
&&\quad ={{\partial ^2} \over {\partial r_1\partial r_2}}\left\langle 0
\right|j(r_1,t_1)j(r_2,t_2)\left| 0 \right\rangle ,
\end{eqnarray}
where $j(r)= (2m)^{-1}\sum_i {[ {p_i\delta (r-r_i)+\delta (r-r_i)p_i} ]}$
is the current operator, we can identify $n(\omega,x)$ with
$\int_{t}^{\infty}\langle j(r,t') j(0,0) \rangle dt'$.
Thus, according to the Kubo formula, Fourier transform of $n(\omega,x)$
with respect to $x^2$ gives the ac conductivity of the one-dimensional
gas.
The correspondence between the universal correlation functions of RMT and
the correlator of CSM is summarized in Table I.



There exist another differential relation between $k(\omega,x)$ and
$n(\omega,x)$ which hold only at $x\to 0$ for $\beta=1$, $2$ and $4$,
namely,
\begin{equation}
  1+k(\omega ,x=0)=\beta \omega ^2{\partial \over {\partial
      x^2}}\left. n(\omega, x )\right|_{x \rightarrow 0}.
\label{second-relation}
\end{equation}
Let us first discuss some physics behind this relation.
The function $n(\omega,x)$ appears when we investigate the universal
properties which depend on the wavefunction as well as the spectrum.  One
of the examples is the dielectric response of a complex crystal with a
chaotic primitive cell, where the quasi-momentum serves as the external
parameter~\cite{Taniguchi93d}.
It is observed that the dynamical conductivity $\sigma(\omega)$ is exactly
proportional to the DOS correlator, $1+k(\omega,0)$~\cite{footnote}.
The statement can be extended to all three ensembles.
This result can be physically explained as the manifestation of
the independence in RMT between the fluctuations of wavefunctions (matrix
elements) and those of the energy spectrum.
However, the  underlying theoretical structure necessary to produce
$\sigma(\omega)\propto 1+k(\omega,0)$ is highly nontrivial, since the
dynamical conductivity should be formally expressed through the
polarization part, hence through $n(\omega,x)$ defined by
Eq.~(\ref{def:n}).
We emphasize that the relation Eq.~(\ref{second-relation}) remains
nontrivial even when we know the explicit integral expression of
$k(\omega,x)$ and $n(\omega,x)$.  In fact, to see this equality explicitly
by starting with the integral expressions such as Eqs.~(\ref{kn-unitary}),
we had to perform all the double- ($\beta=2$) or triple-
($\beta=1,4$) integration by help of the Fourier
transformation.

Gaining insight from Eq.~(\ref{second-relation}), we expect the following
{\em equal-time\/} relation between density-density correlator and
current-current correlator will hold in the Calogero-Sutherland model:
\begin{equation}
  \left\langle 0 \right|j(r)j(0)\left| 0 \right\rangle =-{\lambda
      \over {2m^2 r^2}}\left\langle 0 \right|\rho (r)\rho
    (0)\left| 0 \right\rangle .
\label{current-density}
\end{equation}
We will show below that the identity Eq.~(\ref{current-density}) holds
not only for $\lambda=1/2$, $1$, and $2$, but also for arbitrary values of
$\lambda$, even including {\em irrational\/} values.



To prove Eq.~(\ref{current-density}), we fully exploit the novel structure
of the Calogero-Sutherland model.
The striking feature of CSM hamiltonian is that it is
possessed of the factorization~\cite{Shastry92} and the supersymmetric
structure~\cite{Shastry93}.  When we define the operator
\begin{equation}
 Q_i=p_i+i\lambda
\phi \sum_{j\ne i} {\cot (\phi r_{ij})},
\end{equation}
the Hamiltonian can be factorized as
\begin{equation}
  H={1 \over {2m}}\sum_i {Q_i^{\dagger} Q_i},
\end{equation}
where $[ {Q_i^{\dagger} ,Q_j^{\dagger} }]=[ {Q_i,Q_j}]=0$, and
\begin{eqnarray*}
  \left[ {Q_i,Q_j^{\dagger} } \right] \equiv M_{ij} =
    2\lambda \phi ^2\left[ {\sum_{k\ne i} {\delta_{ij}\over \sin ^{2}(\phi
        r_{ik})}-{(1-\delta _{ij})\over \sin ^{2}(\phi r_{ij})}} \right].\\
\end{eqnarray*}
Note that $Q_i\left| 0 \right\rangle =0$, i.e., $Q_i$ annihilates the
ground state.


To derive the relation Eq.~(\ref{current-density}), first write the current
operator in terms of operator $Q_i$,
\begin{equation}
  j(r)={1 \over {2m}}\sum_i {\left[ {Q_i^{\dagger} \delta (r-r_i)+\delta
      (r-r_i)Q_i} \right]}.
\label{current-Q}
\end{equation}
For simplicity, we will consider the equal-time correlation function
$\langle j(r_1) j(r_2) \rangle$ with $r_1 \ne r_2$.  From
Eq.~(\ref{current-Q}), we get
\widetext
\begin{eqnarray}
  \left\langle 0 \right|j(r_1)j(r_2)\left| 0 \right\rangle && ={1 \over
      {4m^2}}\sum_{ij} {\left\langle 0 \right|\delta (r_1-r_i)\left(
      {M_{ij}+Q_j^{\dagger} Q_i} \right)\delta (r_2-r_j)\left| 0
    \right\rangle },\\ && ={1 \over {4m^2}}\sum_{ij} {\left\langle 0
  \right|M_{ij}\delta (r_1-r_i)\delta (r_2-r_j)\left| 0 \right\rangle }.
\end{eqnarray}
\narrowtext\noindent
As a result, we obtain
\begin{equation}
  \left\langle 0 \right|j(r_1)j(r_2)\left| 0 \right\rangle ={{M(r_{12})}
      \over {4m^2}}\left\langle 0 \right|\rho (r_1)\rho (r_2)\left| 0
  \right\rangle ,
\label{current-density-M}
\end{equation}
where $M(r_{12})=-2\lambda \phi ^2/ \sin ^2(\phi r_{12})$ for $r_1\ne
r_2$.
{}From Eq.~(\ref{current-density-M}), we obtain the equality of
Eq.~(\ref{current-density}) in the thermodynamic limit $r_{12}\ll L$ when
$M(r)\cong -2\lambda / r^2$.

Using Eq.~(\ref{current-density}) and the known expression for $\langle
\rho(r) \rho(0)\rangle$, we get the asymptotic behavior of the
current-current correlator.
\begin{equation}
  \left\langle 0 \right|j(r)j(0)\left| 0 \right\rangle \rightarrow
    \left\{
      \begin{array}{cl}
        {\displaystyle {const. \over {m^2 r^{2-2\lambda}}}}& \mbox{(for $r
          \to +0$)} \\ {\displaystyle {-{\lambda \over {2m^2 r^2}}}}
        &\mbox{(for $r \to \infty$)}
      \end{array}\right.
\end{equation}


We can readily extend this current-density relation to higher point
correlation functions.
\widetext
\begin{eqnarray}
  &&\left\langle 0 \right|j(r_1)\cdots j(r_n)\left| 0 \right\rangle
  = {1\over (2m)^n}\left[ \sum_{\mbox{all possible pairing}}
  {M(r_{\alpha _1}-r_{\alpha _2})\cdots M(r_{\alpha _{n-1}}-r_{\alpha
      _n})} \right]\;\left\langle 0 \right|\rho (r_1)\cdots \rho
  (r_n)\left| 0 \right\rangle.
\end{eqnarray}
For instance, the four-point equal time current correlator is found to be
\begin{eqnarray}
  &&\left\langle 0 \right|j(r_1)j(r_2)j(r_3)j(r_4)\left| 0 \right\rangle
  \nonumber \\ &&\quad ={1\over (2m)^4}\left[
  {M(r_{12})M(r_{34})+M(r_{13})M(r_{24})+M(r_{14})M(r_{23})}
\right]\;\left\langle 0 \right|\rho (r_1)\rho (r_2)\rho (r_3)\rho
(r_4)\left| 0 \right\rangle ,\\ &&\quad \cong {\lambda^2 \over 4m^4} \left
[ {{1 \over
    {(r_{12}r_{34})^2}}+{1 \over {(r_{13}r_{24})^2}}+{1 \over
    {(r_{14}r_{23})^2}}} \right]\;\left\langle 0 \right|\rho (r_1)\rho
(r_2)\rho (r_3)\rho (r_4)\left| 0 \right\rangle .
\label{higher-point}
\end{eqnarray}
\narrowtext \noindent
The last expression is true only for the thermodynamic limit
($r_{ij}\ll L$).
We remark that for $\lambda =1/ 2$, $1$, and $2$, the $n$-point density
correlators were already evaluated by Dyson within the framework of random
matrix theory~\cite{Mehta,Bohigas91}, so Eq.~(\ref{higher-point}) gives us
a way to make an analytical evaluation of the higher-point current
correlators for these values of $\lambda$.  We should substitute the
result for the density correlator in $N\to \infty$ limit which
is presented in the form of the quaternion determinant~\cite{Mehta}:
\begin{equation}
  \left\langle 0 \right|\rho (r_1)\cdots \rho (r_n)\left| 0 \right\rangle
    =\det \left[ {\sigma _\lambda (x_i-x_j)}\right]
\end{equation}
where $1 \le i, j \le n$; $x_i = k_F r_i /\pi$, and
\begin{mathletters}
\begin{eqnarray}
  \sigma _{1/2}(x)&&=\left( {\matrix{{s(x)}&{Ds(x)}\cr
      {Js(x)}&{s(x)}\cr }} \right),\\ \sigma _{1}(x) &&=\left(
  {\matrix{{s(x)}&0\cr 0&{s(x)}\cr }} \right),\\ \sigma _{2}(x) &&=\left
  ( {\matrix{{s(2x)}&{Ds(2x)}\cr {Is(2x)}&{s(2x)}\cr
      }}\right),
\end{eqnarray}
\end{mathletters}
where
\begin{eqnarray*}
&& s(x)={{\sin (\pi x)} \over {\pi x}};\quad
Ds(x)={\partial \over{\partial x}}s(x),\\
&& Is(x)=\int_0^x {{{\sin (\pi x')} \over {\pi x'}}dx'}; \quad
Js(x)=Is(x)-\epsilon(x),
\end{eqnarray*}
where $\epsilon(x)$ is the function which equals to $1/2$, $0$ and $-1/2$,
respectively, for $x>0$, $x=0$ and $x<0$.
%
We note that in a very similar way, we can evaluate equal-time
correlation functions of any combination of current and density operators.

In conclusion, initiating from the relation obtained from the random
matrix theory, we have derived the equal-time relation between the current
correlators and density correlators in the Calogero-Sutherland model,
which can apply to arbitrary values of $\lambda$.  General underlying
structure responsible for this current-density mapping was clarified and
the extension to the higher-point correlation was made.

The authors are grateful to B. D. Simons, E. R. Mucciolo, A. V. Andreev,
and A. Macedo for various useful discussions.
The work of NT was supported in part by the Joint Services Electronic
Program No. DAAL 03-89-0001 and by NSF grant No. DMR 92-14480.

%
%


\begin{references}
\bibitem{footnote} Though the quasimomentum is analogous to the
  Aharonov-Bohm flux and apparently violates T-invariance, this invariance
  is recovered when the primitive cell has the reflection symmetry, as was
  discussed by Robnik and Berry, J. Phys. A {\bf 19}, 669 (1986).
\end{references}

%
\begin{table}[htbp]
\[
\begin{array}{cc}
  \mbox{RMT} & \mbox{CSM}\\
  \hline\hline \\ {\pi \omega} & {k_F r}\\ {\pi ^2x^2/ 2}&{-ik_F^2t}\\
  {1+k(\omega ,x)} & {\left\langle {\rho (r,t)\rho (0,0)} \right\rangle}\\
  {n(\omega ,x)} & {\int_t^\infty {\left\langle {j(r,t')j(0,0)}
    \right\rangle dt'}}\\ \\ \hline\hline
\end{array}
\]
\caption{Correspondence between random matrix theory (RMT) and
  Calogero-Sutherland model (CSM)}
\label{table1}
\end{table}

%

\widetext
\end{document}